\documentclass[twocolumn,showpacs,preprintnumbers,amsmath,amssymb,prb]{revtex4}

\draft

\usepackage{graphicx}
\usepackage{epsf}
\begin{document}

\preprint{CondMat 2004}

\title{Surface Acoustic Waves probe of the $p$-type Si/SiGe heterostructures}

\author{G. O. Andrianov}
\author{I. L. Drichko}
\author{A. M. Diakonov}
\author{I. Yu. Smirnov}
\affiliation{A. F. Ioffe Physicotechnical Institute of RAS, 194021
St.Petersburg, Russia}

\author{O. A. Mironov}
\author{M. Myronov}
\author{T. E. Whall}
\author{D. R. Leadley}
\affiliation{Department of Physics, University of Warwick,
Coventry CV4 7AL, UK}

\date{\today}

\begin{abstract}
{The surface acoustic wave (SAWs) attenuation coefficient $\Gamma$
and the velocity change $\Delta V /V$ were measured for
$p$-type Si/SiGe heterostructures in the temperature range 0.7 -
1.6 K as a function of external magnetic field $H$ up to 7 T and
in the frequency range 30-300 MHz in the hole Si/SiGe
heterostructures. Oscillations of $\Gamma$ (H) and $\Delta V /V$
(H) in a magnetic field were observed. Both real $\sigma_1$ (H)
and imaginary $\sigma_2$ (H) components of the high-frequency
conductivity have been determined. Analysis of the $\sigma_1$ to
$\sigma_2$ ratio allows the carrier localization to be followed as
a function of temperature and magnetic field.  At T=0.7 K the
variation of $\Gamma$, $\Delta V /V$ and $\sigma_1$ with SAW
intensity have been studied and could be attributed to 2DHG
heating by the SAW electric field. The energy relaxation time is
found to be dominated by scattering at the deformation potential
of the acoustic phonons with weak screening.}
\end{abstract}


\maketitle

\section{Introduction}
\label{Introduction}

For the first time an acoustic method has been applied in a study
of $p$-type Si/SiGe heterostructures. Since Ge and Si are not
piezoelectrics the only way to measure acoustoelectric effects in
these systems is a hybrid method: a SAW propagates along the
surface of a piezoelectric LiNbO$_3$ while the Si/SiGe sample is
being slightly pressed onto LiNbO$_3$ surface by means of a
spring. In this case a strain from the SAW is not transmitted to
the sample and it is only the electric field accompanying the SAW
that penetrates into the sample and creates currents that, in
turn, produce a feedback to the SAW. As a result, both SAW
attenuation $\Gamma$ and velocity $V$ appear to depend on the
properties of the 2DHG. SAW-acoustics proves to be an effective
probe of heterostructure parameters, especially as it is {\em
contactless} and does not require the Hall-bar configuration of a
sample. Moreover, simultaneous measurements of attenuation and
velocity of SAW provide a unique possibility to determine the
\emph{complex} AC conductivity,
$\sigma_{xx}(\omega)=\sigma_1(\omega)-i \sigma_2(\omega)$, as a
function of magnetic field $H$ and SAW frequency $\omega$.
Furthermore, the magnetic field dependence of
$\sigma_{xx}(\omega)$ provides information on both the extended
and localized states \cite{ildPRB}.

\section{Experimental Results}

\label{Experiment}

The absorption $\Gamma$ and velocity shift $\Delta V /V$ of the
SAW, that interacts with 2DHG in the SiGe channel, have been
measured at temperatures T=0.7-1.6 K in magnetic fields up to H=7
T. DC-measurements of the resistivity components $\rho_{xx}$ and
$\rho_{xy}$ have also been carried out in magnetic fields up to
11T in the temperature range 0.3-1.3 K and have shown the integer
quantum Hall effect.

The samples were modulation doped Si/SiGe heterostructures with
2DHG sheet density $p = 2\times 10^{11}$ cm$^{-2}$ and mobility
$\mu = 10500$ cm$^2$/Vs \cite{Miron}.

Fig. 1 illustrates the field dependences of $\Gamma$ and $\Delta V
/V$ for the frequency 30MHz at T=0.7 K as well as components of
the magnetoresistance. One can see the absorption coefficient and
the velocity shift both undergo SdH-type oscillations.

\begin{figure}[h]
\centerline{
\includegraphics[width=8cm,clip=]{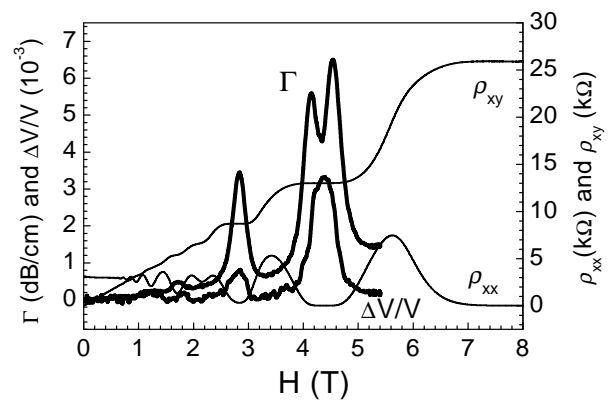}}
\caption{Dependences of $\Gamma$(H) and  $\Delta V /V$(H),
$f$=30MHz, T=0.7 K; $\rho_{xx}$ and $\rho_{xy}$ vs H, T=0.7 K.}
\end{figure}

High frequency conductivity  $\sigma_{xx}^{AC}=
\sigma_{1}-i\sigma_{2}$ is extracted from simultaneous
measurements of $\Gamma$ and  $\Delta V /V$, using eqs. 1-5 of
\cite{ildPRB}.

\begin{figure}[t]
\centerline{
\includegraphics[width=7cm,clip=]{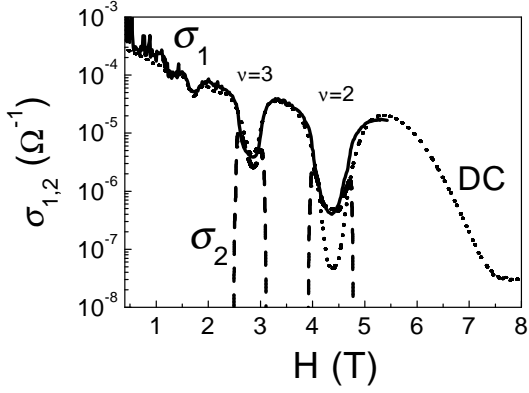}} \caption{Magnetic field
dependences of $\sigma_{1}$ (solid), $\sigma_{2}$ (dashed) at
$f$=30MHz and $\sigma_{xx}^{DC}$ (dotted); all at T=0.7 K.}
\end{figure}

It turns out, that at T=0.7K and filling factor $\nu$=2 (H = 4.3
T) $\sigma_{1}\simeq \sigma_{2}$ (fig.2). At the same time
$\sigma_{1} \gg \sigma_{xx}^{dc}$. These facts suggest that only
some of holes in the 2D-channel are localized, and
$\sigma_{xx}^{AC}$ is determined by both localized and delocalized
holes. For total localization one needs $\sigma_{1}\ll
\sigma_{2}$, $\sigma_{xx}^{DC}$=0 \cite{Efros}. At $\nu$=3 (H=2.9
T) localization effects are negligible: $\sigma_{1} \simeq
\sigma_{xx}^{dc} > \sigma_{2}$.

At T=0.7K we have measured the dependences of $\Gamma$(H), $\Delta
V /V$(H) and $\sigma_{1}$(H) on the SAW intensity at 30MHz. Fig.
3a shows $\sigma_{1}$ versus $P$ (the RF-source power) for
magnetic fields of H = 2.9 T and 4.3 T. Fig.3b illustrates the
temperature dependence of $\sigma_{1}$ measured in the linear
regime. One can see from these plots that $\sigma_{1}$ increases
with increasing temperature and SAW power.

For delocalized holes in this magnetic field, the observed
nonlinear effects (Fig.3a) are probably associated with carrier
heating. One may describe 2DHG heating \cite{MirDeform} by means
of a carrier temperature $T_c$, greater than the lattice
temperature $T$, provided that the following condition is met:

\begin{eqnarray}
\tau_0 << \tau_{cc} << \tau_{\varepsilon}. \label{taurel}
\end{eqnarray}

Here $\tau_0$, $\tau_{cc}$ and $\tau_{\varepsilon}$  are the
momentum relaxation time, the carrier-carrier interaction time and
the energy relaxation time, respectively. Calculations give
$\tau_0= 1.4 \times 10^{-12}$s; $\tau_{cc}$ = 6.4 $\times
10^{-11}$s \cite{FNTMir}; $\tau_{\varepsilon}$ will be discussed
below.

The carrier temperature $T_c$ is determined using SdH thermometry
by comparing the dependences $\sigma_{1}$(T) and $\sigma_{1}$(P).

To characterize heating process one needs to extract the absolute
energy losses as a result of the SAW interaction with the carriers
$\bar{Q}=\sigma_{xx}E^2=4\Gamma W$, where $W$ is the input SAW
power scaled to the width of the sound track, $E$ is the intensity
of the SAW electric field \cite{HeatSAW}:

\begin{figure}[h] \centerline{
\includegraphics[width=7cm,clip=]{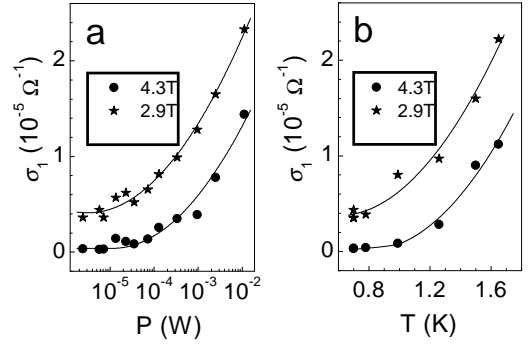}}
\caption{ (a) $\sigma_{1}$ vs the RF-source power $P$;  (b)
$\sigma_{1}$ versus temperature $T$ in the linear regime. $f$=30
MHz, $H$ = 4.3T and 2.9T.}
\end{figure}

\begin{eqnarray}
\  |E|^2=K^2\frac{32\pi}{V}(\varepsilon_1+\varepsilon_0)
\frac{z(q)qe^{(-2qa)}} {1+(\frac{4\pi
\sigma_{xx}^{AC}}{\varepsilon_s V}t(q))^2}W,\label{eq2}
\end{eqnarray}
$K^2$ is the electromechanical coupling constant of the lithium
niobate (Y-cut), $q$ is the SAW wave vector; $\varepsilon_s$,
$\varepsilon_0$ and $\varepsilon_1$ are the dielectric constants
of semiconductor, free space and LiNbO$_3$, respectively; $a$ is
the width of the sample-LiNbO$_3$ clearance; $z(q)$ and $t(q)$ are
functions to allow for the electrical and geometrical properties
of the sample.

We have analyzed the energy losses rate per hole $Q=\bar{Q}/p$ as
a function of $T_c$ (Fig.4). In the inset, $Q$ is plotted vs $
(T_c^5-T^5)$ and a linear dependence is illustrated.

\begin{figure}[h] \centerline{
\includegraphics[width=5cm,clip=]{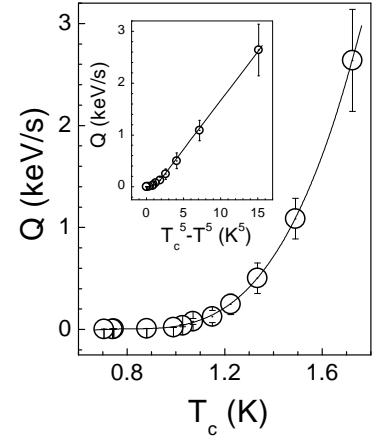}}
\caption{The energy losses rate per hole $Q=\bar{Q}/p$ plotted vs
$T_c$. Inset: dependence of $Q$ on ($T_c^5-T^5$).}
\end{figure}

For weak heating ($\Delta T=T_c-T\ll T$) one can estimate the
energy relaxation time from $\tau_{\varepsilon}=(\pi
k_B)^2/(3\gamma A_{\gamma} \varepsilon_F T^{\gamma -2})$
\cite{HeatSAW}, where $\gamma$=5 in this case, $\varepsilon_F$ is
the Fermi energy and $A_5$ is the slope of the $Q$($T_c^5-T^5$)
dependence. Using this equation we have computed
$\tau_{\varepsilon} =(3.8 \pm 0.4) \times 10^{-8}$s. Thus,
relation (\ref{taurel}) is satisfied.

The observed dependence of the energy loss rate on the carrier
temperature as $Q=A_5 (T_c^5-T^5)$ corresponds to the case of
energy relaxation via carrier scattering from the deformation
potential of the acoustic phonons with weak screening
\cite{Karpus, MirDeform}. In which case the slope $A_5$ is
\cite{Karpus}:

\begin{eqnarray}
A_5=\frac{3 \sqrt{2} m^2 \zeta(5) D^2_{ac} k^5_B}{\pi^{5/2} s^4
\hbar^7 p^{3/2} \rho} ,
\end{eqnarray}
where $\rho$ is the mass density, $s$ is the longitudinal sound
velocity, $m$=0.24$m_e$ is the effective mass \cite{FNTMir}. Thus,
one can determine the value of the deformation potential as
$D_{ac}$=5.3$\pm$0.3 eV. The value of $D_{ac}$ calculated from DC
measurements of phonon-drag thermopower was reported to be
5.5$\pm$0.5 eV for the same 2DHG Si/SiGe sample \cite{MirAgan}.

\acknowledgments The work was supported by RFFI, NATO-CLG 979355,
INTAS-01-084, Prg. MinNauki "Spintronika".

\end{document}